\documentclass[journal=aaemcq,manuscript=letter]{achemso}

\usepackage[version=3]{mhchem} 
\usepackage{bm}
\usepackage{amsmath}
\usepackage{subcaption}
\usepackage{graphicx}
\usepackage{textcomp}
\usepackage{multirow}
\usepackage{array}
\usepackage{tabularx}
\usepackage{booktabs}  
\usepackage{caption}
\usepackage[table]{xcolor}
\usepackage{hyperref}

\author{Vinay Thakur}
\author{Prabhat Prakash}
\email{pprakash@caltech.edu}
\altaffiliation{Chemistry and Chemical Engineering, California Institute of Technology, Pasadena, California, 91125 United States}
\author{Raghavan Ranganathan}
\email{rraghav@iitgn.ac.in}
\affiliation{Department of Materials Engineering, Indian Institute of Technology Gandhinagar, Gujarat, 382355, India}

\title[An \textsf{achemso} demo]
  {How individual vs shared coordination governs the degree of correlation in rotational vs residence times in a high-viscosity lithium electrolyte}

\begin{document}

\newpage

\begin{abstract}
Commercially used carbonate-based electrolytes in lithium-ion batteries are susceptible to many challenges, including flammability, volatility, and lower thermal stability compared to \textcolor{black}{LiTFSI-glyme based electrolytes}. Mixtures of LiTFSI salt (lithium bis(trifluoromethylsulfonyl)-amide) and glyme-based solvents are potential alternative candidates for commonly used electrolytes. We perform classical molecular dynamics (MD) simulations \textcolor{black}{to} study the effect of concentration and temperature on the translational and rotational dynamics. The radial distribution function shows stronger coordination of Li$^+$ ions with tetraglyme(G4), as shown in earlier studies, and forms a stable [Li(G4)]$^+$ cation complex. The self-diffusion coefficients are lower than the values experimentally observed but show improvement over other classical force fields, without charge scaling. An increase in the salt concentrations leads to a higher viscosity of the system and reduces the overall ionic mobility of Li$^{+}$ ions. Diluting the system with a larger number of G4 molecules leads to shorter rotational relaxation times for both TFSI and G4. Ion-residence times show that Li$^+$ ions form stable and long-lasting complexes with G4 molecules than TFSI anions. The residence time of [Li(G4)]$^+$ complex increases in highly concentrated system due to the availability of fewer G4 molecules to coordinate with a Li$^+$ ion. G4 is also seen to form polydentate complexes with Li$^+$ ions without a shared coordination, allowing rotation without breaking coordination, unlike TFSI, which requires coordination disruption for rotation. This distinction explains the poor correlation between rotation and residence time for G4 and the strong correlation for TFSI.
\end{abstract}

\newpage
\section{Introduction}

The growing demand for clean energy across the world requires efficient energy storage devices. Batteries and supercapacitors are among the widely used electrochemical devices for energy storage. Lithium-ion batteries (LIBs) have made significant progress in the past three decades in reducing their overall cost while increasing energy density. The cost has decreased from \$1,000 per kWh in 2009 to \$180 per kWh in 2020, while the energy density has increased from 125 Wh/kg to 250 Wh/kg. The LIB market was \$36.7 billion in 2019 and is expected to reach \$128.3 billion by 2027, with an estimated compound annual growth rate of 18\% from 2020 to 2027.\cite{feng2022review,nitta2015li}. The electrolyte is an important component of batteries that determines the cyclic efficiency and the charge-discharge rate of a battery. Generally, ideal electrolytes should possess high ionic conductivity and good electrochemical and thermal stability. Commercialized LIBs use electrolytes of fluoride salts and organic solvents, which provide high ionic conductivity but are susceptible to leakage and flammability and are responsible for most fire accidents in electric vehicles. Various alternative electrolyte systems, such as ionic liquids (ILs),\cite{RANA2023116340} solvate ionic liquids (SILs),\cite{Nachaki2024} polymer electrolytes,\cite{wu2024fire}localized highly concentrated electrolytes,\cite{van2023comparative} inorganic solid electrolytes,\cite{Yu2025} soft-solid electrolytes,\cite{Prakash2023} and composite electrolytes,\cite{ZHANG2023100316} have been proposed to increase the safety aspects and energy density of the batteries. SILs are considered an alternative to conventional organic electrolytes owing to their unique structural characteristics and higher ionic conductivity. SILs generally form in equimolar mixtures when a cation (e.g., Li$^+$ here) forms a coordinated complex ([Li(G4)]$^+$) with the donor sites of the solvent molecule, with anions being solvent-separated and occupying the interstitial volume. SILs have higher viscosity compared to organic electrolytes, still they possess higher ionic conductivity which is  attributed to the higher concentration of Li(G4)[TFSI].\cite{tamura2010physicochemical,di2022glyme} The glyme family of solvents, specifically those with higher chain lengths, are a better alternative to organic solvents in ILs due to their lower flammability, less toxicity, and higher thermal stability. In ILs, the currently used higher chain length solvents are triglyme (G3), tetraglyme (G4), and pentaglyme (G5) with LiTFSI and LiPF$_6$ salts. LiTFSI salt shows excellent thermal and chemical stability up to 360 \textdegree C and dissociates easily in various types of solvents.\cite{wang2019review,DiLecce2022,lingua2024solvate,rajendran2024electrochemical,dong2018charge}

Higher chain-length glyme molecules have better physiochemical properties compared to shorter chain-length glyme molecules. Equimolar concentrations of triglyme (G3) and tetraglyme (G4) have been investigated with lithium  salts using various experimental (Raman spectroscopy, Small angle X-Ray Scattering, Diffusion nuclear magnetic resonance)  and computational techniques.\cite{aguilera2015structural,yoshida2011change,tamura2010physicochemical} Structural properties like radial distribution function (RDF) and radius of gyration ($R_\text{g}$) are useful in predicting the local solvation shell and compactness of atoms or molecules, respectively. The local structure of Li$^+$ ions prefers to remain in a five-fold (or four-fold) coordination with oxygen of G4 and TFSI combined, where the major contribution comes from G4.\cite{tsuzuki2014structures,callsen2017solvation,sun2018insight} Based on G4 or TFSI coordination, the LiTFSI/G4 complexes are classified as contact ion pairs (CIP) and solvent-separated ion pairs (SSIP). In CIP, oxygen atoms on TFSI directly coordinate with solvated Li$^+$ ions, whereas in SSIP, TFSI anions interact with the [Li(G4)]$^+$ complex. The SSIP configurations were found to be energetically more favorable compared to CIP, stating that Li$^+$ ions prefer to remain as a [Li(G4)]$^+$ complex in the system, also supported by the diffusivity ratio ($D_{\text{Li}^+}/D_{\text{G4}}$), which is equal to one for equimolar systems.\cite{aguilera2015structural,zhang2014chelate,ho2023understanding}

Experimental techniques like 2D IR spectroscopy, along with molecular mechanics-based non-reactive classical molecular dynamics (MD) simulations and first principle calculations, have been used to assess and characterize the distribution of these various structural complexes.\cite{rajput2015coupling,callsen2017solvation,rushing2019effect,galvez2021ion,chen2024dominant,tan2024decoding} However, the challenge with classical MD is that it does not capture the structural changes in solvated structures during the dynamics very well compared to Ab initio (AIMD) simulations. Still, inaccessibility to longer timescales in AIMD simulations forces the use of classical models, especially for longer time-correlated properties like translational and rotational diffusion. Recently, machine learning interatomic potentials (MLIPs) have emerged as an efficient alternative to both classical and ab initio molecular dynamics for liquid electrolyte simulations in batteries. These models can achieve near–first-principles accuracy while dealing with large system sizes and longer timescales accessible to classical MD.\cite{magduau2023machine,goodwin2024transferability,gong2025predictive,levine2025open}

Experimentally, the diffusion coefficients for Li$^+$ , TFSI, and G4 are calculated at 30 \textdegree C and 80 \textdegree C (discussed later in the subsection \textit{Translational dynamics}).\cite{tamura2010physicochemical,yoshida2011change,zhang2014chelate,harte2023accelerated} Using MD simulations with a polarizable force field, Dong et al.\cite{dong2018charge} found that the diffusion coefficients of species at 30 \textdegree C and 80 \textdegree C are one or two order of magnitude less than the experimental diffusion coefficients. Other research groups also used non-polarizable OPLS-force fields with scaled charges (to include polarization effects) and found similar diffusion coefficients at 227 \textdegree C.\cite{thum2020solvate} No experimental study has been done so far  that highlights the system dynamics at such elevated temperatures\cite{harte2023accelerated}. The use of scaled charges can model the structure right by over-simplification of the many-body interactions. Thus, the use of more robust methods to compute the partial charge of molecules is required. Dong's group also calculated the lifetime of the Li$^+$ ions with TFSI and G4 at 373 K, which depends upon the salt concentration. A smaller concentration of salt leads to a greater number of free glyme molecules in the system, which goes through a faster exchange event with glyme molecules binding with  Li$^+$ ions, thus reducing the lifetime of the $\text{Li-O}_{\text{G4}}$.\cite{dong2018charge} The translation diffusion, exchange events, and residence times also dictate the coiling and bending of geometries of G4 and TFSI, respectively. However, no modeling of rotational dynamics has been conducted yet that shows the effect of the coiling/bending of one molecule on another or on the rotational lifetime of molecules. The rotational contribution is an important factor in deciding the lifetime of big continuous clusters of Li$^+$ ions with TFSI and G4. Moreover, the effect of correlated motions on the ionic mobility is not very well understood even in some of the recent studies,\cite{mitra2024exploring,philippi2025,bhabani2025} which motivates us to investigate if there is an interplay between residence and rotational lifetimes in such high-viscosity lithium electrolytes. Although, the absolute timescales might be overestimated due to simulation limitations, the observed relative trends and the correlation between residence and rotational times are expected to be reliable and physically meaningful.

\begin{figure}[ht!]
    \centering
    \includegraphics[width=0.95\linewidth]{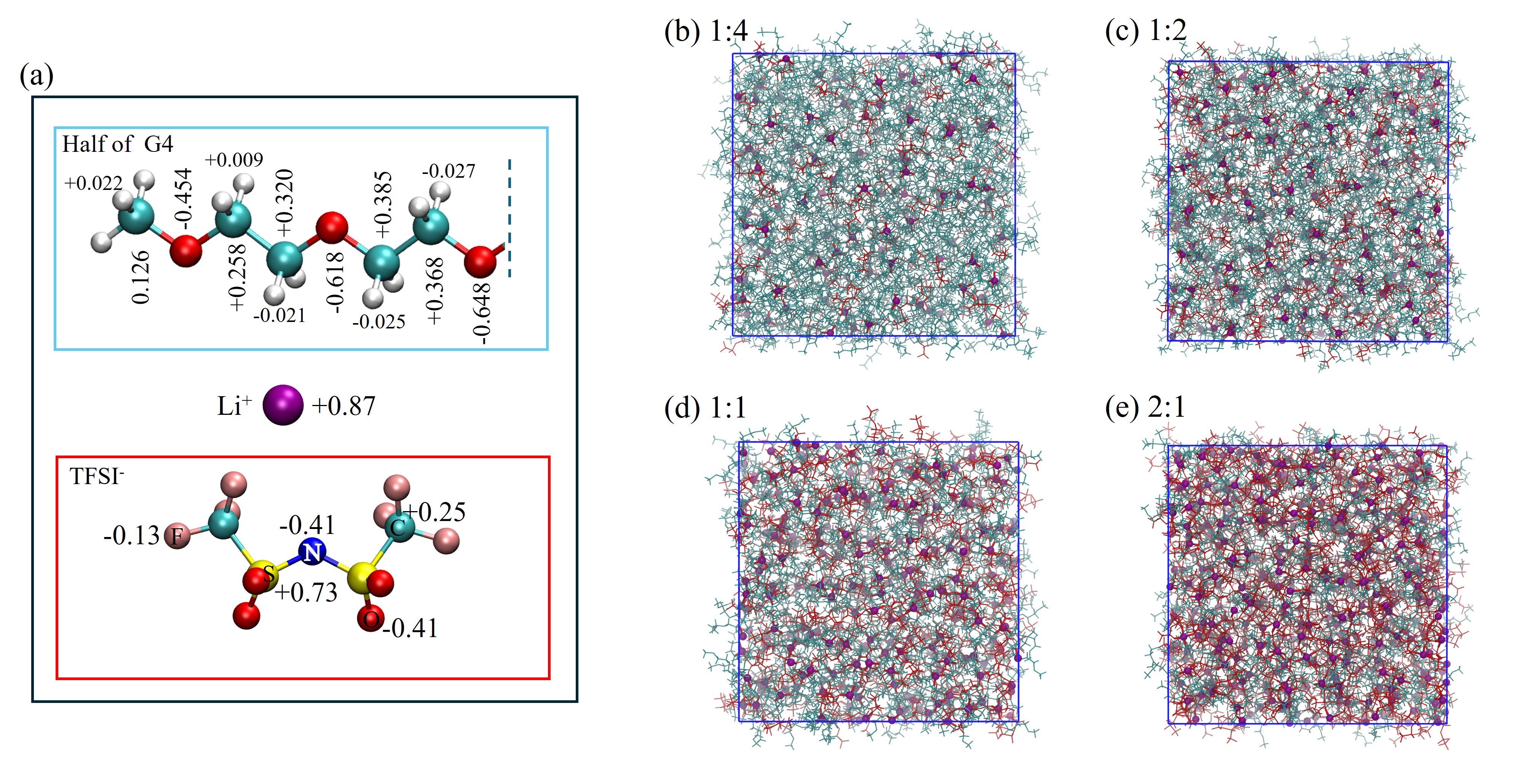}
    \caption{(a) Schematic representation of LiTFSI and G4 with partial atomic charges, (b) 1:4, (c) 1:2, (d) 1:1 and (e) 2:1 ratio mixtures of LiTFSI/G4 at 300 K after an equilibration of 10 ns.}
    \label{fig:str_chg}
\end{figure}

The present work investigates the extreme range of LiTFSI and G4 mixture compositions (from 1:4 to 2:1) at various temperatures with classical MD simulations using a modified OPLS force-field.\cite{jorgensen1996development} The partial charges for atoms of LiTFSI and G4 are reparametrized here (Figure \ref{fig:str_chg}). A correct structure prediction and long timescale simulations performed in the present work provide a reliable set of results for diffusion coefficients, residence times, rotational relaxation times, and cluster sizes and theorize a not-yet-explored interplay of these.

\section{Computational Details}

Four mixtures with different LiTFSI/G4 ratios (1:4, 1:2, 1:1, and 2:1) were studied, as shown in Table S1 of the Supporting Information (SI). These mixtures were modeled using a charge-reparameterized OPLS force field. Partial atomic charges for the LiTFSI and G4 molecules were calculated (Figure\ref{fig:str_chg}a) using a range-separated hybrid functional wB97x-D4 functional\cite{wb97,GD4} with a sufficiently large basis set def2-tzvpp, utilizing the ORCA 5.0.1 package\cite{ORCA5}. Partial atomic charges were derived for all atoms of TFSI and G4 from their geometries in Li-bonded states using CHELPG method\cite{chelpg}. A fixed charge of +0.87 was assigned to Li$^{+}$ ions based on the bonded state of the cation and free cation. The PME\cite{essmann1995smooth} algorithm was used to compute the long-range electrostatic interactions. The cut-off for Lennard-Jones and electrostatic interactions is set to 10 Å. A time-step of 1 femtosecond is used throughout the simulations. All simulations were conducted using the Gromacs 2021 \cite{gromacs} software package. For post-processing and analysis, VMD 1.9.3 \cite{humphrey1996vmd}, Travis\cite{brehm2020travis}, and in-house scripts developed in python were used.

The different ratio mixtures were initially minimized using the conjugate-gradient algorithm. Each system was heated from 300 K to 600 K in the NPT ensemble with a heating rate of 50 K per 5 ns to mix it well, followed by an equilibration run in the NVT ensemble for 10 ns at each particular temperature. Figure \ref{fig:str_chg}(b-e) shows post-equilibrated mixtures at 300 K. Velocity rescaling method\cite{bussi2007canonical}  and Berendsen barostat\cite{berendsen1984molecular} were used to manage fluctuations in temperature and pressure, respectively.  Following this, the systems were cooled again from 600 K to 300 K in the NPT ensemble at the interval of 50 K per 5 ns with 10 ns of NVT equilibration at each particular temperature. Final production runs were performed at 300 K and 500 K in NVT ensemble conditions for 100 ns to obtain sufficiently long trajectories for analysis.

\section{Results and discussion}

\subsection{Solvation structure}

\begin{figure}[ht!]
    \centering
    \begin{subfigure}{0.49\textwidth}
    \subcaption*{\textbf{(a)}}\label{rdf:a}
    \centering
    \includegraphics[width=1.0\textwidth]{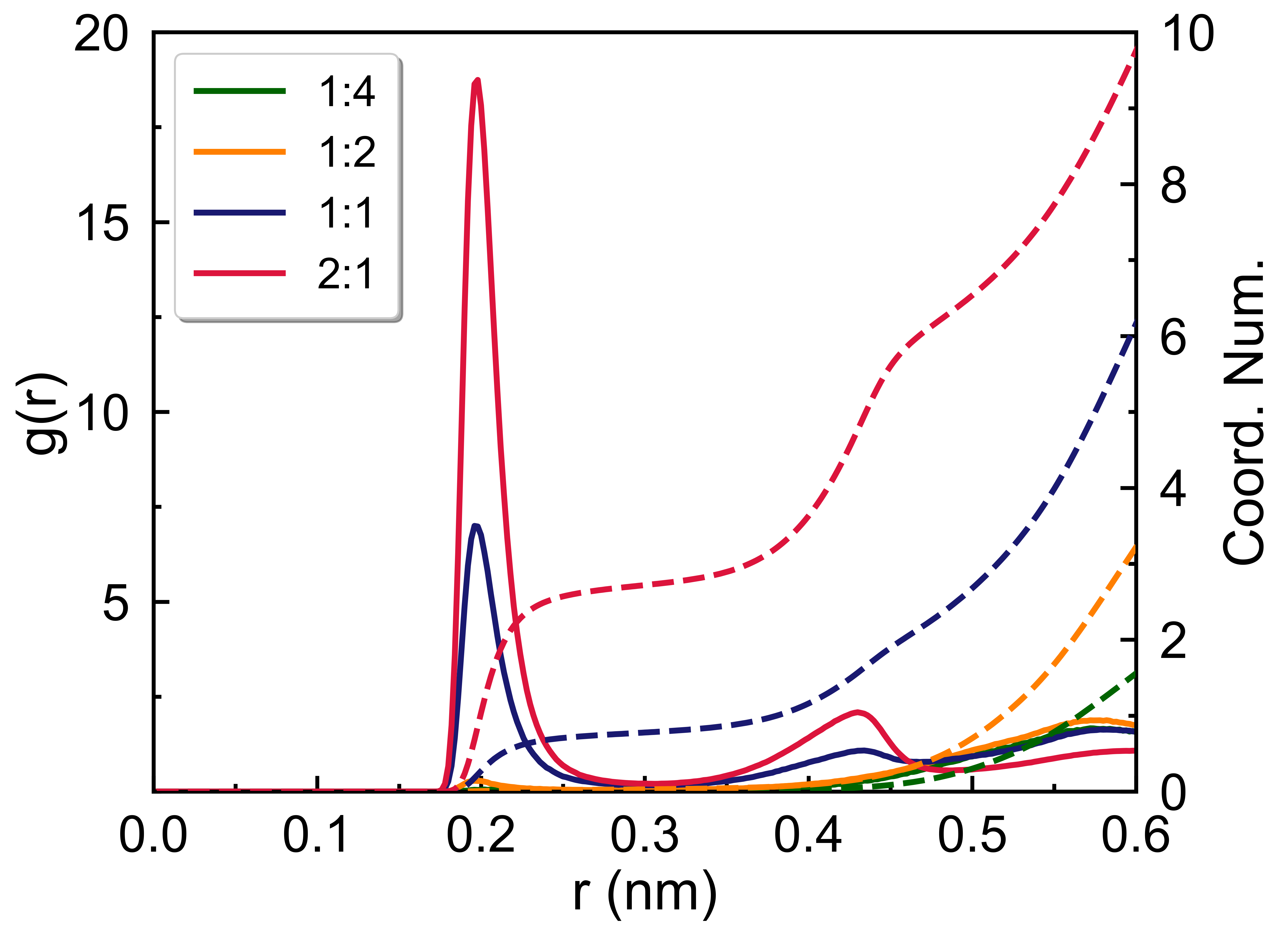} 
    \end{subfigure}
    \hfill
    \begin{subfigure}{0.49\textwidth}
    \subcaption*{\textbf{(b)}}\label{rdf:b}
    \centering
    \includegraphics[width=1.0\textwidth]{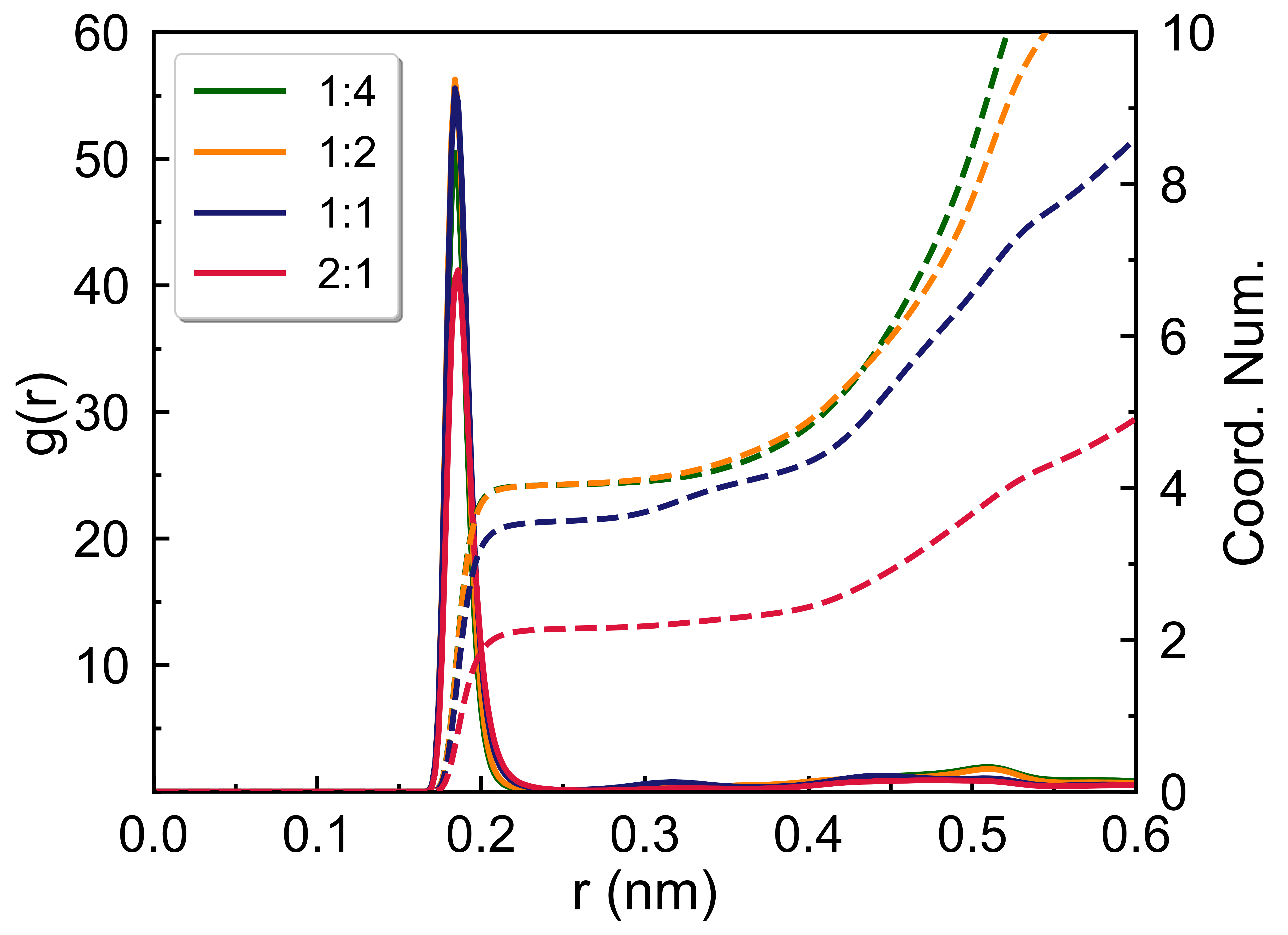}
    \end{subfigure}
    \hfill
    \begin{subfigure}{0.49\textwidth}
    \subcaption*{\textbf{(c)}}\label{rdf:c}
    \centering
    \includegraphics[width=1.0\textwidth]{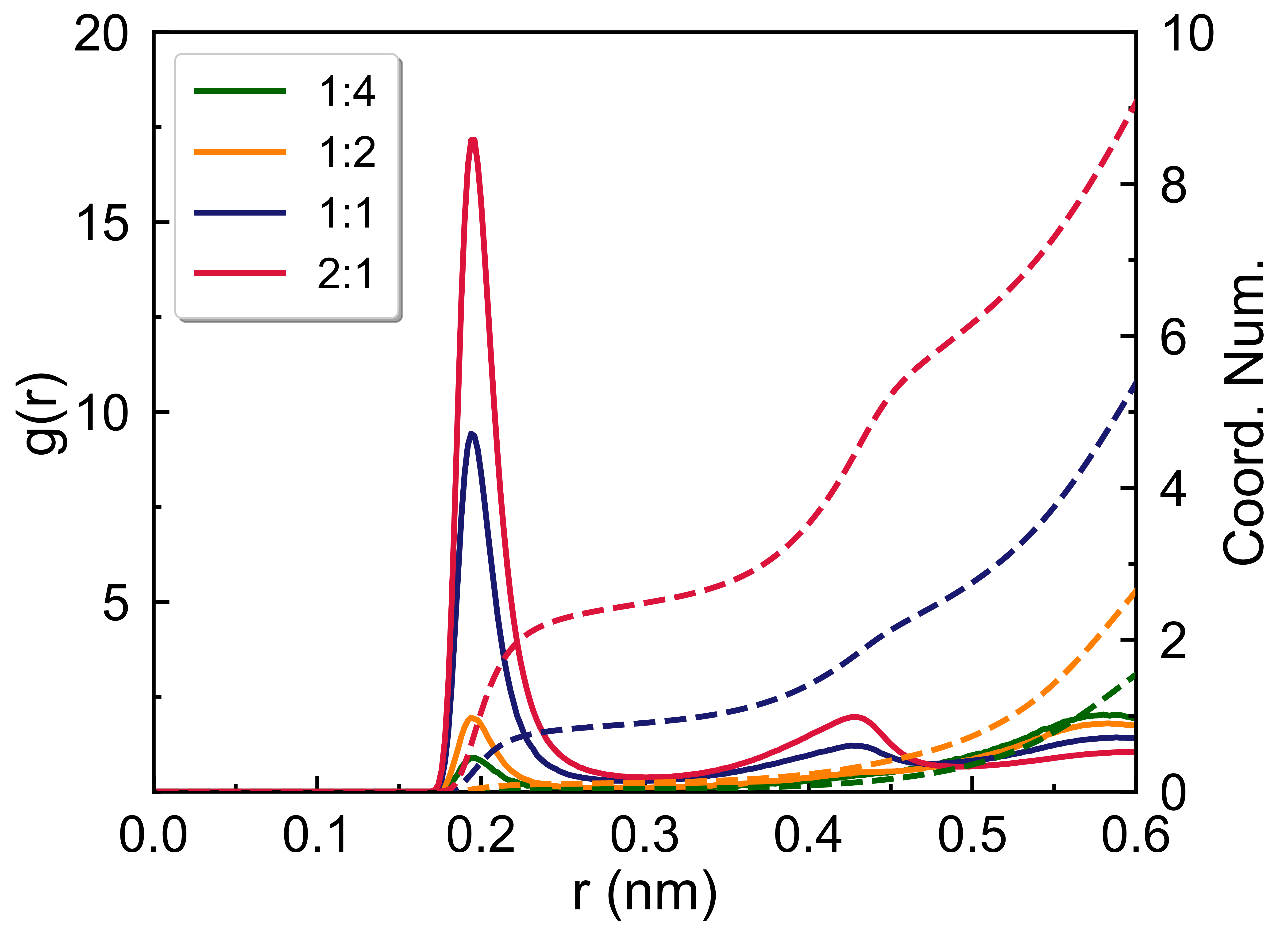}
    \end{subfigure}
    \hfill
    \begin{subfigure}{0.49\textwidth}
    \subcaption*{\textbf{(d)}}\label{rdf:d}
    \centering
    \includegraphics[width=1.0\textwidth]{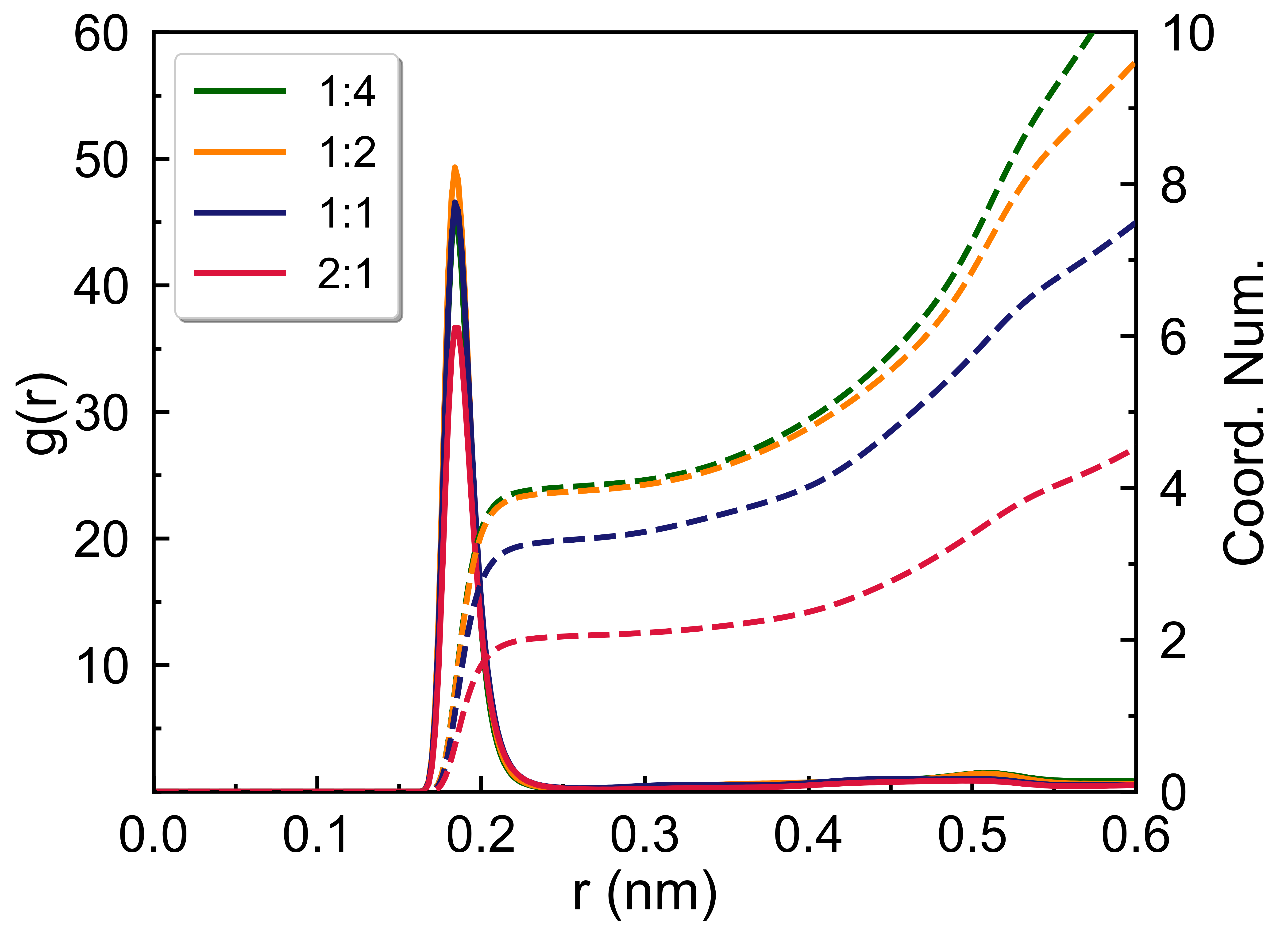}
    \end{subfigure}

    \vspace{5pt} 

\caption {Radial distribution function for (a) Li...O(TFSI), (b) Li...O(G4) at 300 K, and (c) Li...O(TFSI), (d) Li...O(G4) at 500 K for different compositions.}
    
    \label{fig:rdf_plots}
\end{figure}

The interatomic interactions of TFSI and G4 around Li$^+$ ions are calculated as radial distribution functions (RDF):

\begin{equation}
g(r)=\frac{n(r)}{\rho 4 \pi r^{2} \Delta r}
\end{equation}
and coordination number, 
\begin{equation}
n(r^{'})=4\pi\rho\int_{0}^{r^{'}}g(r)r^{2}\,dr\
\end{equation}

$n(r)$ is the number of particles located at the distance between $r$ and $r+\Delta r$ from the reference particle, $\rho$ is the bulk number density, and $r^{'}$ is the distance of the first minima of $g(r)$ from each atom. The neutron scattering confirms that in an equimolar mixture of LiTFSI and G4, 95 \% G4 molecules are coordinated by Li$^+$ ions and the remaining 5 \% have an incomplete Li...O(G4) coordination.\cite{murphy2016bulk} The experimental $g(r)$ peaks are observed at 1.9 \text{\AA} for Li...O(G4), and 2.0 \text{\AA} for Li...O(TFSI) in an equimolar (1:1) mixture.\cite{murphy2016bulk} Our simulations are in a good agreement with these observations as the first solvation shell of a Li$^+$ ion is seen to form at 1.9 \text{\AA} with O atoms of G4 and at 2.0 \text{\AA} with O atoms of TFSI (Figure \ref{fig:rdf_plots}). From the integration of these RDFs (within a cut-off radius of 2.3 \text{\AA}), the coordination number in the equimolar ratio at 300 K is 3.6 for Li...O(G4) and $\simeq$ 0.7 for Li...O(TFSI). The variation in the coordination number of Li$^+$ ions is observed at different salt concentrations. With the increase in the concentration of glyme molecules, the coordination number of Li...O(G4) increases nearly to 4 at 1:2 and 1:4 ratios. Also, the increase in the concentration of solvent leads to an increase in the intensity of RDF of the Li...O(G4), whereas the intensity of RDF of Li...O(TFSI) decreases as more number of Li$^+$ ions find free G4 and get attached to them. Changing the concentration of salt and solvent does not affect the radius of the first coordination shell around lithium.

The coordination number of Li...O(G4) in equimolar ratio decreased to 3.3 (500 K) from 3.6 (300 K). This could happen due to the accelerated dynamics, which leads to a faster exchange between G4 molecules at higher temperatures within the solvation shell of lithium. This also leads to a decrease in the lifetime of Li...O(G4) complex which is confirmed through ion-pair/ion-solvent existence autocorrelation function (discussed \textit{vide infra} in subsection \textit{Ion-residence times}). Also, the intensity of RDF peaks showed a small decrease at 500 K for both Li...O(TFSI) and Li...O(G4) when compared to RDF at 300 K. Additional Li...Li RDFs are provided in Table S1 of SI.

\subsection{Radius of gyration}

Since both the TFSI anion and G4 are long molecular structures, their coiling leads to a significant variation in the radii of gyration ($R_\text{g}$). We used MD trajectories to calculate $R_\text{g}$ for TFSI anions and G4 molecules for each ratio and at 300 K and 500 K (Figure S2 and S3 in SI). As shown in Table \ref{tab:gyration_table}, the $R_\text{g}$ of TFSI and G4 in the most diluted system (1:4) at 300 K is, 0.256 nm and 0.381 nm, respectively. At a higher concentration, 2:1, the $R_\text{g}$ for both TFSI and G4 decreases to 0.254 nm and 0.363 nm, suggesting a more coiled configuration, particularly for G4, which shrinks by ~5 \%. The $R_\text{g}$ for both TFSI and G4 decrease in highly concentrated system. This indicates that at higher salt concentrations, most G4 molecules coordinate individually with the Li$^+$ ions, thus increasing their overall compactness. In diluted system, there are more free G4 molecules available to share the coordination leading to higher $R_\text{g}$. As expected at higher temperature, $R_\text{g}$ for TFSI and G4 increase from 0.256 nm to 0.257 nm and from 0.381 nm to 0.387 nm respectively, when increasing the temperature from 300 K to 500 K. Even these slight changes in $R_\text{g}$ is an important factor which can lead to an interplay in translational and rotational dynamics and ion-residence times in these mixtures (discussed \textit{vide infra}). 

\begin{table}[H]
    \centering
    \renewcommand{\arraystretch}{1.0}
        \caption{Average and standard deviation of radius of gyration, $R_\text{g}$(nm),  for 1:4 and 2:1 system}
    \begin{tabular}{>{\centering\arraybackslash} m{2.2cm} >{\centering\arraybackslash} m{2.2cm} >{\centering\arraybackslash}  m{2.2cm} >{\centering\arraybackslash}  m{2.2cm} >{\centering\arraybackslash}  m{2.2cm} >{\centering\arraybackslash}  m{2.2cm} }

        & & \multicolumn{2}{c}{\textbf{Temp (300 K)}} & \multicolumn{2}{c}{\textbf{Temp (500 K)}} \\
         \cmidrule(lr){3-4} \cmidrule(lr){5-6}
          & \textbf{Species} & \textbf{Average} & \textbf{Std. dev.} &  \textbf{Average} & \textbf{Std. dev.}\\
    
        \multirow{2}{*}{1:4} & TFSI & 0.256 & 0.0006 & 0.257 & 0.0006 \\
         & G4 & 0.381 & 0.0016 & 0.389 & 0.0018 \\  
        \multirow{2}{*}{2:1} & TFSI & 0.254 & 0.0002 & 0.255 & 0.0003 \\  
         & G4 & 0.363 & 0.0007 & 0.367 & 0.0034 \\  

    \end{tabular}

    \label{tab:gyration_table}
\end{table}

\subsection{Translational dynamics}

\begin{figure}[t!]
    \centering
    \includegraphics[width=0.95\linewidth]{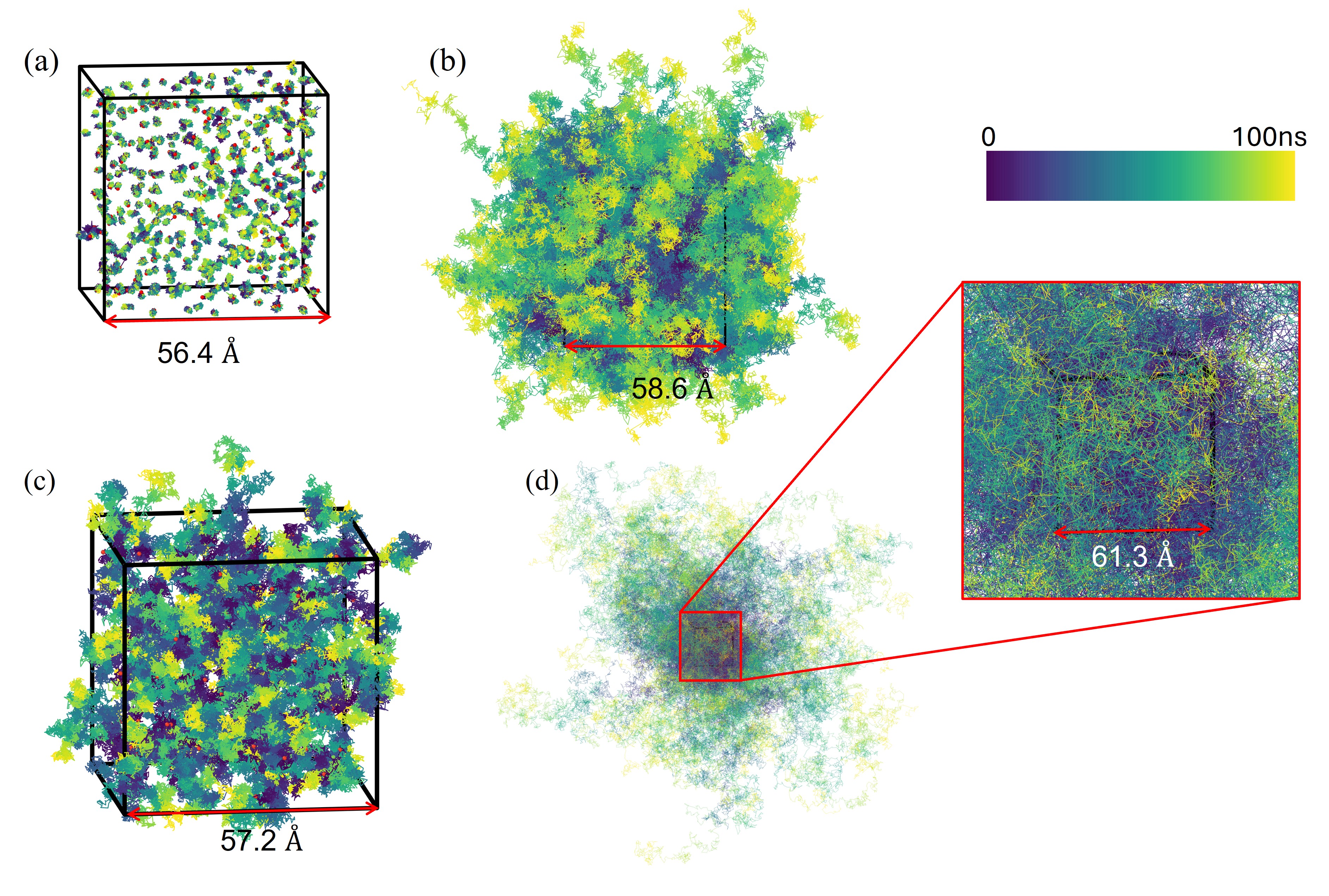}
    \caption{Trajectory of Li$^+$ ions over a time of 100 ns for different LiTFSI/G4 ratios (a) 2:1 at 300 K, (b) 2:1 at 500 K, (c) 1:4 at 300 K, (d) 1:4 at 500 K.}
    \label{fig:traj_Li}
\end{figure}

\begin{table}
\centering
\renewcommand{\arraystretch}{1.0}

\caption{Comparison of self-diffusion coefficients $D$ (in $10^{-7}$ cm$^{2}$/s) for equimolar (1:1) LiTFSI/G4 electrolyte from experiments and MD simulations.}

\begin{tabular}{l l l l}
\textbf{Temp (K)} & \textbf{$D_{\mathrm{Li^+}}$} & \textbf{$D_{\mathrm{TFSI^-}}$} & \textbf{$D_{\mathrm{G4}}$} \\
\hline

\multicolumn{4}{c}{Experiments$^{a}$ (Tamura et al.)} \\
303 & 1.26 & 1.22 & 1.26 \\

\multicolumn{4}{c}{Experiments$^{a}$ (Yoshida et al.)} \\
303 & 1.31 & 1.22 & 1.29 \\

\multicolumn{4}{c}{Experiments$^{a}$ (Zhang et al.)} \\
303 & 1.26 & 1.22 & 1.26 \\

\multicolumn{4}{c}{Experiments$^{a}$ (Harte et al.)} \\
303 & 1.01 & 1.04 & 0.87 \\
353 & 27.90 & 7.08 & 6.33 \\

\multicolumn{4}{c}{MD$^{b}$ (Dong et al.)} \\
303 & 0.41 & 0.40 & 0.41 \\

\multicolumn{4}{c}{MD$^{c}$ (Tsuzuki et al.)} \\
403 & 0.78 & 0.73 & 0.78 \\

\multicolumn{4}{c}{MD$^{c}$ (Shinoda et al.)} \\
503 & 4.5 & 5.7 & 4.6 \\

\multicolumn{4}{c}{MD$^{d}$ (Thum et al.)} \\
500 & 53 & 59 & 54 \\

\multicolumn{4}{c}{MD$^{e}$ (Harte et al.)} \\
500 & 52.02 ± 1.06 & 56.63 ± 3.23 & 53.10 ± 1.46 \\
700 & 96.16 ± 16.02 & 113.12 ± 6.56 & 96.69 ± 10.69 \\

\multicolumn{4}{c}{MD$^{f}$ (This work)} \\
300 & 0.02 ± 0.01 & 0.04 & 0.03 ± 0.01 \\
500 & 12.43 ± 0.27 & 23.7 ± 2.51 & 12.38 ± 0.02 \\
\hline
\end{tabular}

\vspace{0.2cm}
\footnotesize
\textbf{$^{a}$} PFG-NMR. \\
\textbf{$^{b}$} MD simulation, polarizable APPLE\&P force field. \\
\textbf{$^{c}$} MD simulation, OPLS-AA and CL\&P force field, no charge scaling. \\
\textbf{$^{d}$} MD simulation, OPLS-AA and CL\&P force field, with scaled charges. \\
\textbf{$^{e}$} MD simulation, OPLS-AA force field, with scaled charges. \\
\textbf{$^{f}$} MD simulation, OPLS-AA force field, no charge scaling. \\

\label{tab:diffcoeff}
\end{table}

The translational diffusion of ions and G4 is ascertained with the calculation of mean square displacement (MSD) for all trajectories. Since  Li$^+$ ions are strongly coordinated with O atoms of TFSI and G4, at high-concentration (e.g., 2:1) and low-temperature (e.g. 300 K), their movement (Li$^+$ ions) is heavily vibrational in their coordination ‘cage’. The trajectory lines for 2:1 and 1:4 compositions (Figure \ref{fig:traj_Li}) at 300 K and 500 K show that for 2:1 at 300 K (Figure 3a), all Li$^+$ ions move in the coordination cage with a vibrational motion of 0.1 - 0.3 Å of amplitude between two frames separated over 5 ps time-interval. These cage-vibrations are also observed in other inorganic and soft-solid electrolytes\cite{Prakash2023,three-salts2025}. The self-diffusion coefficients ($D$) are obtained from the diffusive regime of MSD vs time plots (Figure S4 and S5 in SI), utilizing the Einstein relationship as:

\begin{equation}
      D= \frac{1}{2d\Delta t}\frac{1}{N}\sum_{i=1}^{N}\langle (|r_i(t+\Delta t)-r_i(t)|^{2})\rangle
\end{equation}
\begin{figure}[h!]
    \centering
    \includegraphics[width=0.5\linewidth]{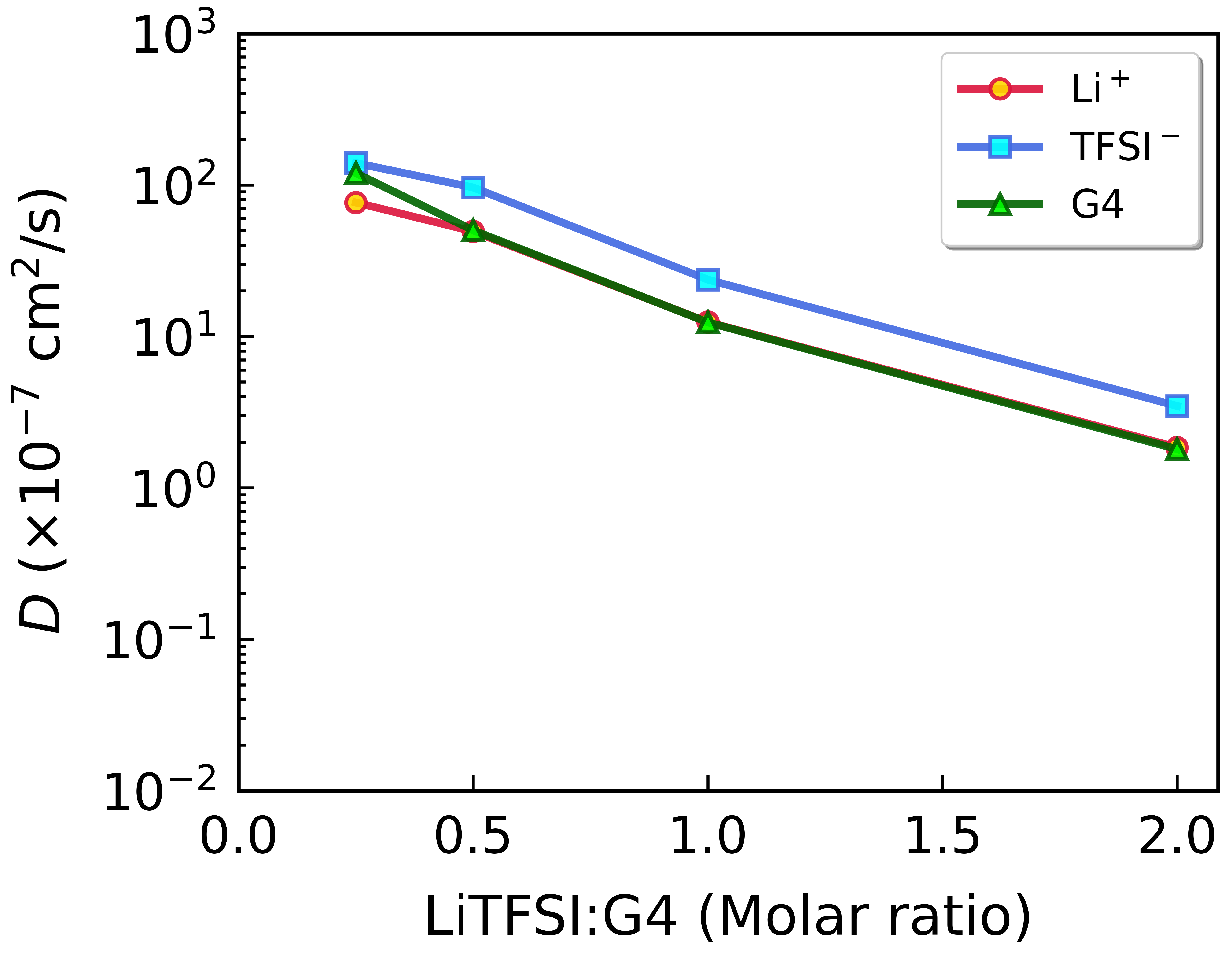}
    \label{fig:diff-coeff}
\caption{Self-diffusion coefficients of Li$^+$, TFSI$^-$, and G4 for different LiTFSI/G4 ratios at 500 K}
\label{fig:log_diffusion}
\end{figure}

Here, $r_{i}(t)$ is the position of $i$th atom at time $t$, $\Delta t$ is the lag time, $d$ is the dimensionality of the system, $N$ is the number of particles, and $\langle (|r_i(t+\Delta t)-r_i(t)|^{2})\rangle$ is the mean-squared displacement averaged over multiple origins of time. The calculated values of self-diffusion coefficients at 300 K and 500 K are reported in Table S2 of SI, and are provided in Figure \ref{fig:log_diffusion} (at 500 K). Experimentally, the self-diffusion coefficient of Li$^+$ ions ($D_{\ce{Li^+}}$) is measured using both impedance spectroscopy and pulsed-field gradient NMR (Table \ref{tab:diffcoeff}). \cite{yoshida2011change,zhang2014chelate, harte2023accelerated} In simulations, classical MD with scaled charges (a brute-force way to obtain the appropriate local interactions and diffusion) predicts self-diffusion coefficients better than the other non-scaled charge-based works.\cite{thum2020solvate, harte2023accelerated} Here, the diffusion coefficients of G4 are similar to those of Li$^+$ ions for a 1:1 system, showing the formation of stable [Li(G4)]$^+$ complexes. This suggests that lithium primarily diffuse in "vehicular-type" mechanism, in the form of [Li(G4)]$^{+}$ complex\cite{zhang2014chelate,ho2023understanding}. With more G4 molecules available in the mixture, $D_{\ce{Li^+}}$, $D_{\ce{TFSI^-}}$ and $D_{\ce{G4}}$ increase due to faster exchanges between G4 molecules inside Li$^+$ ion solvation shell. Li...Li RDFs  (Figure S1 in SI) also show that the most probable distance between two Li$^+$ ions is 4-6 \AA, similar to what is observed in highly concentrated solid electrolytes\cite{Prakash2023,paul-lii}. This includes the possibility of hopping between transient coordination sites along with vehicular motion, leading to sluggish diffusion of Li$^+$ ions. Also, increasing the concentration of LiTFSI leads to an increase in the pairing probability of [Li(G4)]$^+$ and TFSI anions, causing sluggish diffusion of both ions.\cite{sun2018insight} The calculated transference numbers for Li$^+$ are close to 0.34 (Table S3), considering no anion-blocking, thus are closer to PFG-NMR value (\ $\sim$ 0.5) while significantly overestimated compared to potentiostatic polarization estimates (0.02 - 0.03)\cite{tli-1, tli-2}.  

\subsection{Rotational dynamics}

The rotational autocorrelation function (RACF) is employed to analyze the orientational dynamics of TFSI and G4. In the case of TFSI anion, the S-N-S triplet is selected, where the rotational vector is defined as the cross-product of the two S-N and N-S bond vectors. Similarly, for the G4 molecule, the atom triplet consists of C-O-C atoms, where O is the geometric center of the molecule. The RACF for the vector is written as:

\begin{equation}
 C(t)=\langle (u(0)\cdot u(t) \rangle   
\end{equation}

where $u(0)$ and $u(t)$ are the rotational vectors $u$ at time t = $0$ and t = $t$. 

Figure S6 in SI shows that for the system with the highest concentration of G4 molecules (1:4), RACF for both TFSI and G4 decays slower with increasing salt concentration. In dilute mixtures, larger number of free G4 molecules in the system are partially/not coordinated with Li$^+$ ions, resulting in fewer obstructions in their rotational dynamics. Also, viscosity in LiTFSI-based electrolytes rises with increasing LiTFSI concentration but decreases significantly with the temperature rise, with reports indicating a reduction of up to 83 \% in few cases and more than that, over a 50 K temperature rise.\cite{Saito2017a,Zhang2021} With increasing salt concentration, the rotational decay becomes much slower, due to the higher viscosity of the system and increase in the shared coordination of TFSI and G4 with Li$^+$ ions. Also, G4 molecules rotate more slowly than TFSI anions, primarily due to a bigger radius of gyration and also due to their higher coordination with Li$^+$ ions. 

The rotational relaxation time $\tau_{rot}$ (Table \ref{tab:rot_time}) is calculated as:

\begin{equation}
   \tau_{rot} = \int_{0}^{\infty} C(t)\,dt\
\end{equation}

At 300 K, the C(t) for some compositions did not decay sufficiently during the simulation time; hence $\tau_{rot}$ for those is calculated analytically here by fitting C(t) with two exponential functions, with the fitting coefficients shown in Table S4 and S5 in SI. At 300 K, $\tau_{rot}$ is 295 ns for TFSI and 893 ns for G4 for 2:1 mixture. As the amount of G4 increases, $\tau_{rot}$ decreases sharply for both TFSI and G4. $\tau_{\text{rot}}$ at 300 K is in the order of ns which decreases to ps as the temperature is raised to 500 K. The local structure and the nature of coordination of Li$^+$ ions alters  $\tau_{rot}$ for a couple of orders across these concentrations. An analysis on the correlation of $\tau_{rot}$ with the residence time of G4...Li and TFSI...Li and formation of interatomic networks (clusters) is discussed further to understand this complexity. 

\begin{table}[h!]
    \centering
    \renewcommand{\arraystretch}{1.0}
        \caption{Rotational relaxation time, $\tau_{\text{rot}}$(ns), for TFSI and G4 for different concentrations at 300 K and 500 K}
    \begin{tabular}{>{\centering\arraybackslash} m{2.9cm} >{\centering\arraybackslash}  m{2.9cm} >{\centering\arraybackslash}  m{2.9cm} >{\centering\arraybackslash}  m{2.9cm} >{\centering\arraybackslash}  m{2.9cm} } & \multicolumn{2}{c}{\textbf{Temp (300 K)}} & \multicolumn{2}{c}{\textbf{Temp (500 K)}} \\

         \cmidrule(lr){2-3} \cmidrule(lr){4-5}

        & $\boldsymbol{\tau}_{\textbf{rot,TFSI}}$ & $\boldsymbol{\tau}_{\textbf{rot,G4}}$ & $\boldsymbol{\tau}_{\textbf{rot,TFSI}}$ & $\boldsymbol{\tau}_{\textbf{rot,G4}}$ \\

         1:4 & 0.2 & 6.02 & 0.005 & 0.03 \\     
         1:2 & 0.59 & 22.59 & 0.005 & 0.06 \\  
         1:1 & 15.7 & 189.15 & 0.020  & 0.16 \\     
         2:1 & 295.22 & 893.31 & 0.260 & 0.59 \\      
    \end{tabular}

    \label{tab:rot_time}
\end{table}

\subsection{Ion-residence times}

The residence time of Li$^+$ aggregates formed with TFSI and G4 is computed as an ion-pair (or ion-solvent) existence autocorrelation function (IEAF), which can be written as an auxiliary function

\begin{equation}
\beta_{ij}(t) =
\begin{cases} 
1, & \text{TFSI and G4 inside Li$^+$ solvation shell} \\
0, & \text{otherwise}
\end{cases}
\end{equation}

and the IEAF for $\beta_{ij}$ is

\begin{equation}
 C(t)= \frac{1}{N^2}\sum_{i=1}^{N}\sum_{j=1}^{N}\int_{0}^{\infty}\beta_{ij}(t)\cdot\beta_{ij}(t+\tau)\,dt\
\end{equation}

 The residence time $\tau_{res}$ of the aggregates formed can be computed as:

\begin{equation}
     \tau_{res}= 2\int_{0}^{\infty} C(t)\,dt\
\end{equation}

$\tau_{res}$ for Li...O(G4) and Li...O(TFSI) interactions is calculated within the first solvation shell of lithium. The radius of the solvation shell is taken from the first minima of the corresponding RDFs. Figure S7 in SI shows the decay of IEAF for Li-O, for both TFSI and G4 at 300 K and 500 K. The decay is too sluggish for Li...O(TFSI) and Li...O(G4) in highly concentrated system. For all these systems  $\tau_{res}$ is calculated numerically (for well-decayed) and analytically (for sluggish-decayed) tabulated in Table S6 in SI. The value of $\tau_{res}$ is  $\sim$ 6 $\mu$s for Li...TFSI and $\sim$ 30 $\mu$s for Li...G4 for 2:1 mixture at 300 K, which decreases to $\sim$ 1 ns and $\sim$ 882 ns, respectively, as the system is diluted with more G4 (1:4). Dong et al.\cite{dong2018charge} calculated the residence time of Li...O(TFSI) at various temperatures and showed that, for the 1:1 system, the residence time for Li...O(TFSI) is $\sim$ 7 ns at 303 K and decreases with increasing the concentration of free G4 molecules and at higher temperatures. In our study, for equimolar system, the residence time of Li...O(TFSI) is 445 ns at 300 K and decreased to 1 ns at 500 K, as shown in Table S6 in SI. Similarly, the residence time of Li...O(G4) is $\sim$ 34 $\mu$s at 300 K and decreases by three orders to $\sim$ 46 ns at 500 K. The discrepancy in the residence time between our study and Dong's group could be due to the differences in force fields and the treatment of partial charges. They used polarizable force field that capture the dynamic molecular charge redistribution in response to the changing environment, which in turn weakens the coordination, facilitates exchange, and consequently reduces the residence time. In a similar study for an equimolar LITFSI/G4, Shinoda et al.\cite{shinoda2018molecular} reported that the residence time of Li...TFSI and Li...G4 at 503 K was 2.5 ns and 87 ns, respectively, which is closure to the values we observed. \textcolor{black}{The slight differences may be due to differences in how partial charges were assigned}. More dilution leads to an increase in the number of free G4 molecules in the system, enabling their rapid exchange with the Li$^+$ ion coordination shell, subsequently decreasing the residence time. 

\begin{figure}[h!]
    \centering
    \includegraphics[width=1.0\linewidth]{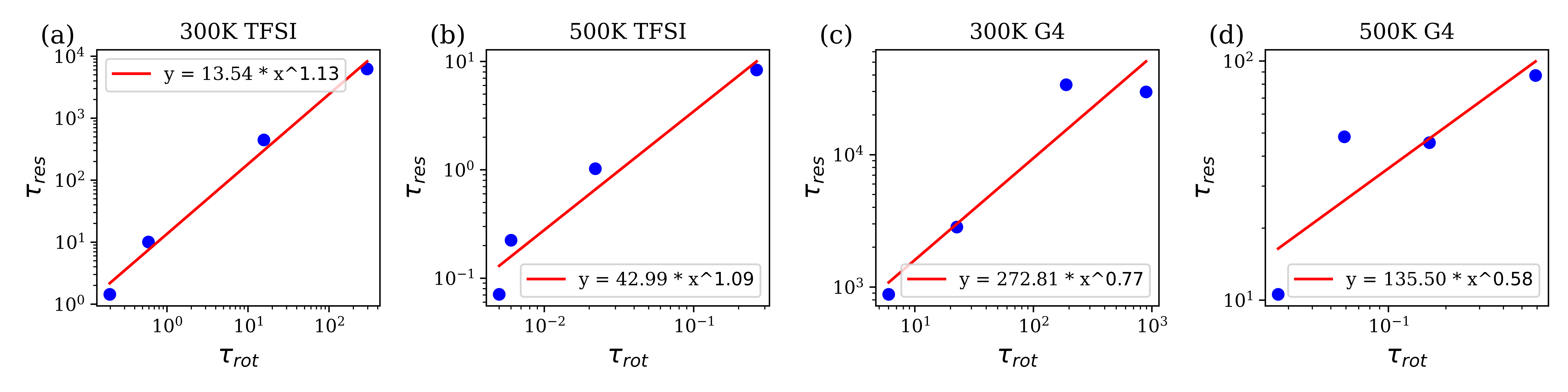}

\caption{Correlation plots for $\tau_{res}$ vs. $\tau_{rot}$ of (a) TFSI at 300 K, (b) TFSI at 500 K, (c) G4 at 300 K, (d) G4 at 500 K.}
\label{fig:corr_res_rot}
\end{figure}

We propose that such long-residence times, particularly at high concentration and low temperature, could affect the rotational motion of these entities with a significantly large $R_g$. If an ion-solvent or ion-pair network has a long lifetime ($\tau_{res}$), then their $\tau_{rot}$ would also be proportionately longer. To probe this, we computed the correlation of $\tau_{res}$ and $\tau_{rot}$ for each G4 and TFSI, at 300 K and 500 K. Here each dataset has four pairs of values for the different concentrations of G4 and TFSI. Plotted on a log-log scale (Figure \ref{fig:corr_res_rot}), lifetimes (rotational-individual and residence with Li) of TFSI anion are fitted as $x^{1.13}$ and $x^{1.09}$ at 300 K and 500 K, respectively - exhibiting a greater than 99\% correlation in both cases of temperature. However, for G4, there is surprisingly low-correlation of 66\% (at 300 K) and 89\% (at 500 K) between the lifetimes. It can be understood that an increase from 300 K to 500 K in temperature simply aligns the rotational motion with the residence time better, but there is a stark contrast of this correlation with TFSI. To investigate further, we analyzed ionic aggregates, most-probable clusters and their nature below. 

\subsection{Ionic aggregates and clusters}

\begin{figure}[h!]
    \centering
    \includegraphics[width=1.0\linewidth]{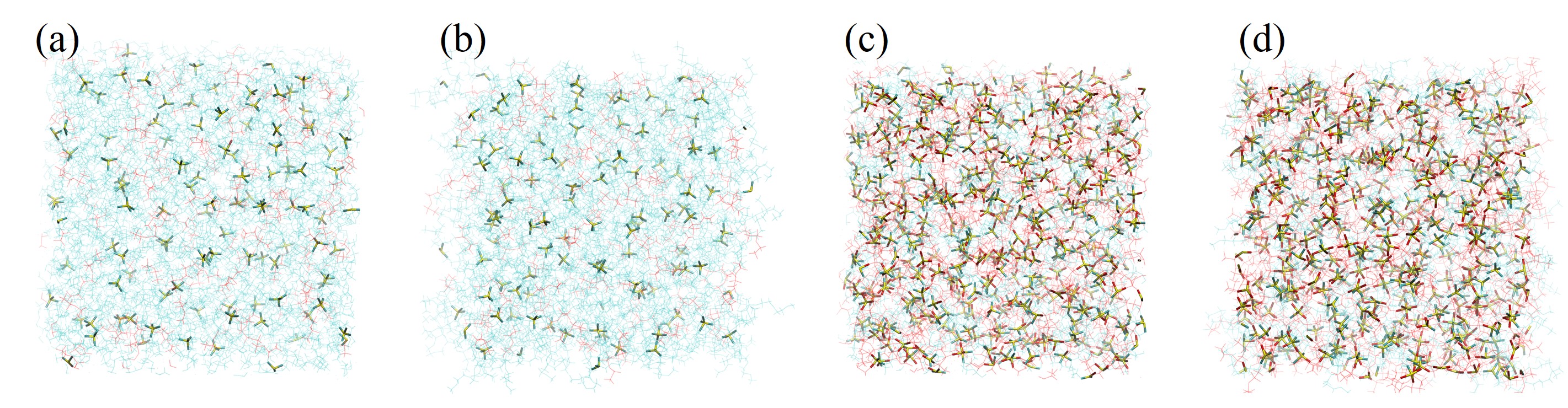}

\caption{Dynamic bonds $\leq$ 2.3 Å showing Li...O(G4) (yellow-green) and Li...O(TFSI) (yellow-red) clusters in LiTFSI/G4 molar ratios (a) 1:4 at 300 K, (b) 1:4 at 500 K, (c) 2:1 at 300 K, and (d) 2:1 at 500 K. Thick lines, yellow Li, red O(TFSI), green O(G4); Thin lines, red TFSI, green G4.}

\label{fig:dynamic_bonds}
\end{figure}

\begin{table}[h!]
\centering
\caption{Number of uncoordinated molecules of TFSI or G4 and the size of the biggest exclusive clusters of {Li...TFSI} and {Li...G4} at {300K} and {500K}.}
\begin{tabularx}{\textwidth}{>{\centering\arraybackslash}X >{\centering\arraybackslash}X >{\centering\arraybackslash}X >{\centering\arraybackslash}X >{\centering\arraybackslash}X >{\centering\arraybackslash}X >{\centering\arraybackslash}X >{\centering\arraybackslash}X >{\centering\arraybackslash}X}
\toprule
 & \multicolumn{4}{c}{\textbf{300K}} & \multicolumn{4}{c}{\textbf{500K}} \\
\midrule
 & \multicolumn{2}{c}{\textbf{Li-TFSI}} & \multicolumn{2}{c}{\textbf{Li-G4}} & \multicolumn{2}{c}{\textbf{Li-TFSI}} & \multicolumn{2}{c}{\textbf{Li-G4}} \\
\midrule
 & \textbf{Uncoord.} & \textbf{Size of Max.} & \textbf{Uncoord.} & \textbf{Size of Max.} & \textbf{Uncoord.} & \textbf{Size of Max.} & \textbf{Uncoord.} & \textbf{Size of Max.} \\
\midrule
1:4 & 5.7	&	187.3	&	0.1	&	539.9	&	5.3	&	172.5	&	0.4	&	539.6 \\
1:2 & 7.6	&	415.1	&	1.7	&	645.6	&	8.4	&	412.2	&	2.3	&	644.6 \\
1:1 & 3  	&	397	    &	0.4	&	455.5	&	2.8	&	436 	&	0.9	&	327.9 \\
2:1 & 11.5	&	852.5	&	4	&	631.1	&	8.6	&	845.2	&	3	&	637.9
 \\
\bottomrule
\end{tabularx}
\label{tab:separete_cluster}
\end{table}

\begin{table}[ht!]
    \centering
    \renewcommand{\arraystretch}{1.0}
    \caption{Number of uncoordinated molecules (TFSI and/or G4) and size of the biggest shared cluster involving Li..G4, Li...TFSI or G4...Li...TFSI.}
    \begin{tabular}{>{\centering\arraybackslash} m{2.5cm} >{\centering\arraybackslash}  m{2.5cm} >{\centering\arraybackslash}  m{3.0cm} >{\centering\arraybackslash}  m{2.5cm} >{\centering\arraybackslash}  m{2.5cm} }
    
         & \multicolumn{2}{c} {\textbf{Temp (300 K)}} & \multicolumn{2}{c} {\textbf{Temp (500 K)}} \\ 

        \cmidrule(lr){2-3} \cmidrule(lr){4-5}
       
       & \textbf{Single} & \textbf{Maximum} & \textbf{Single} & \textbf{Maximum}  \\
       
1:4 & 77.6	&	569 	&	77.8	&	553 \\
1:2 & 147.8	&	712	    &	122.9	&	705 \\
1:1 & 91.5	&	637  	&	65.2	&	634 \\
2:1 & 5.9	&	1070	&	9.6	   &	1062
    \end{tabular}
   \label{tab:whole_cluster}

\end{table}

\begin{figure}[ht!]
    \centering
    \includegraphics[width=1.0\linewidth]{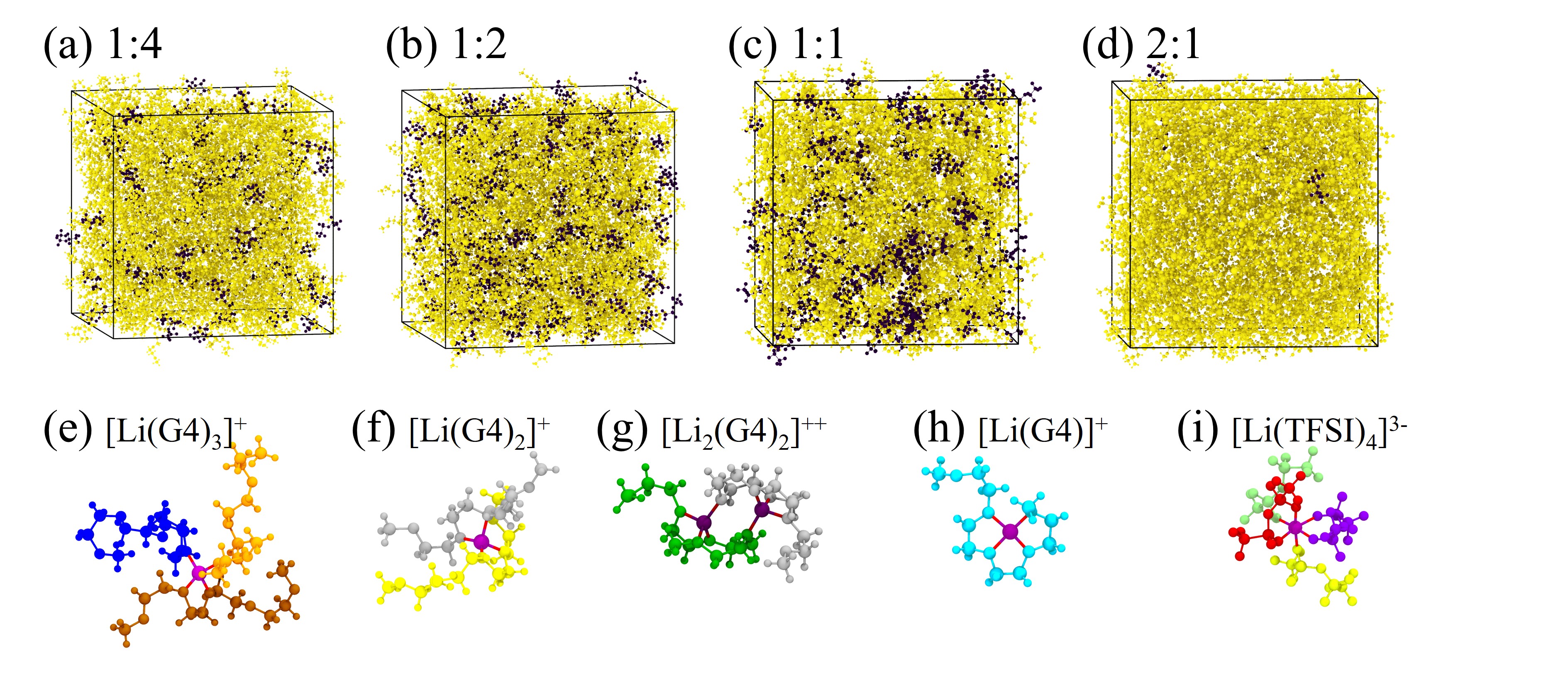}

\caption{Single largest cluster (in yellow) and all other small clusters (in black) for (a) 1:4, (b) 1:2, (c) 1:1 and (d) 2:1 LITFSI/G4 at 300 K, the snapshots show the uncoordinated molecules (black), in each case, which decrease with concentration; Most probable finite clusters from a trajectory of 90 ns, (e), (f), (g), (h), and (i).}
    \label{fig:clust-size}
\end{figure}

All the concentrations of LITFSI/G4 are observed to form long networks where Li, TFSI and G4 coordinate in a shared and unshared manner to form various clusters. Figure \ref{fig:dynamic_bonds} shows the dynamic bonds around each Li$^+$ ion in each case of concentration at 300 K. The snapshots show that for the dilute mixtures (1:4), most of these clusters are individual, and/or unshared (Figure \ref{fig:dynamic_bonds}
a, b). For the concentrated mixtures at 2:1 ratio, long-chains, shared coordination lead to the formation of complex aggregates (Figure \ref{fig:dynamic_bonds} c, d). The aggregation of molecules as long vs. short networks is analyzed by calculating the number and size of clusters during the production run for last 90 ns of simulation time. A cut off of 2.3 \text{\AA}, chosen from RDF, is set as the criteria for cluster formation. We define a cluster when one or more atoms of a G4 or TFSI are within 2.3 \text{\AA} to a Li$^+$ ion. We analyzed these networks as two different ways - individual and shared. For individual networks, mutually exclusive Li...TFSI and Li...G4 clusters are calculated (Table \ref{tab:separete_cluster}). For shared networks, every possible cluster involving Li or G4 or TFSI within 2.3 \text{\AA} are calculated (Table \ref{tab:whole_cluster}). For these networks, we look at the two extreme conditions : the size of the single largest network and the number of smallest cluster (of size 1, which is essentially an uncoordinated G4 or TFSI).  For 1:4 mixture at 300 K, the number of molecules that do not form any cluster is nearly equal to 6 (out of 108) which are mostly TFSI anions. This suggests that in case of excessive solvation with G4, the free uncoordinated molecules are TFSI. For concentrated systems (2:1), $\sim$11 (out of 432) TFSI and 4 (out of 216) G4 molecules are left uncoordinated. The analysis of all (individual+shared) clusters shows that  $\sim$14\% TFSI or G4 (total 77.6) are uncoordinated and free at 1:4 at 300 K. As the system is more concentrated, this number first increases for 1:2 to 22\% and then decreases for more concentrated systems, 18\% for 1:1, and less than 1\% for 2:1. 
The largest networks at various concentrations, which involve the majority of TFSI, G4 forming these via a shared coordination with Li$^+$ ions, are shown in Figure \ref{fig:clust-size} (a - d). All intermediate sized, and intermittent shared clusters  with their trajectory averaged occurrences are provided as size-dependent distribution histograms in Figure S8. Since most of the molecules/ions are part of either the single largest cluster or the uncoordinated single molecule/ions. The distribution histogram shows that the intermediate sized clusters increase with G4 in the system, and at high-temperature also, which leads to formation of small aggregates of size 2 - 5 (long-lived) and 6 - 20 (short-lived). 

We obtained some of the prominent finite geometries from these systems (Figure \ref{fig:clust-size} e - i), which unravel the difference in the degree of correlation in $\tau_{res}$ and $\tau_{rot}$ of TFSI vs. G4. The key arguments are as follows with the index of subfigures: \newline
(e) For 1:4 system, shared clusters mostly involving multiple G4 around a Li$^+$ ion are formed; since only one or rarely two O(G4) coordinate with a Li$^+$ ion, the rest O(G4) coordinate with other Li$^+$ ions and thus forming a long ...(G4)..Li...(G4)...Li... network, leaving no individual G4 uncoordinated. 
\newline
(f) For 1:2 system, some G4 bidentate with a Li$^+$ ion; the other O(G4) atoms form extended networks similar to 1:4.  \newline
(g) For 1:1 system, some Li$^+$ ions form close-contact aggregates, which have two Li$^+$ ions bidented by two G4 molecules. \newline
(h) shows single unshared complex of G4 with a Li$^+$ ions; such complexes rarely form shared networks, except at high concentrations ($\geq$ 1:1), where anions dentate additionally raising the coordination number from 4 to 5 or 6. 
\newline
(i) shows a Li...TFSI cluster which forms only via shared coordination from multiple TFSI. A single TFSI cannot form a complex like (h) due to rigidity and other geometrical constraints.  

It is important to notice here that G4 can form individual unshared complexes with a Li$^+$ ion in a polydentate manner, which TFSI cannot. Thus, a rotation of Li...G4 complex can occur without the need of breaking the coordination, while in case of Li...TFSI, break in coordination is essential for any rotation of an individual TFSI. This explains the poor correlation between $\tau_{rot}$ and $\tau_{res}$ for G4, and highly correlated those for TFSI. A longer residence time only restricts the rotation when there is coordination from multiple similar entities, where the diffusion of Li$^+$ ions can have a significant contribution from hopping.  A polydentate anion or solvent coordinating in an unshared manner would not affect its rotation with its residence time and thus can live longer while rotating faster. This core finding was also validated using scaled charge-based force-field parameters from Harte et al.,\cite{harte2023accelerated} which led to improved diffusion coefficients at 500 K, D$_{Li}^+$ = 54.8 $\times$ 10$^{-7}$ cm²/s (vs. Harte et al.: 52.02 $\times$ 10$^{-7}$ cm²/s) (Figure S9). The central hypothesis also remains valid, as the correlations in ion-residence vs. rotational relaxation are seen preserved, almost independent of charge scaling effect (99\% correlation for TFSI and 84–89\% correlation for G4 between residence and rotational times). The additional rotational and ion-existence correlation functions and the correlation in these lifetimes are provided in Figure S10 and S11. In future, the use of machine-learning interatomic potentials trained on AIMD data, and inclusion of a polarizable charge equilibration, although expensive, could be useful in capturing the system dynamics more precisely. The correlation observed here is potentially an important descriptor, that can be further tested in extending the general theories for concentrated electrolytes\cite{swan2021,mundy2024} to those with long-chain entities.

\section{Conclusion}
This work demonstrates the impact of varying salt concentrations and temperature on the structure, dynamics and interplay of residence vs. rotational relaxation times in LiTFSI/G4 mixtures with classical MD simulations. The structural analysis shows that lithium ions preferentially coordinate with glyme molecules, forming [Li(G4)]$^+$ complexes, complementing the earlier experimental and computational studies. Temperature, in the range of 300 K to 500 K, did not significantly affect the local solvation structure of Li$^+$ ion, suggesting that the coordination environment remains stable even at higher temperatures. The dynamic properties indicated that the self-diffusion coefficients of Li$^+$ ions and solvent molecules decreased with increasing salt concentration, resulting in a sluggish movement of ions. This suggests that higher salt concentrations increase the viscosity of the system, thus restricting the overall ionic mobility. In highly concentrated system, a larger network of clusters is formed, which causes sluggish dynamics of the ions and molecules in the system. 

The ability of G4 to form individual unshared polydentate complexes with Li$^+$ allows rotation without breaking coordination, while TFSI must disrupt coordination to rotate. This difference accounts for the varying correlations between rotation and residence time in both cases. A longer residence time restricts rotation only when multiple similar entities coordinate, while a polydentate species with unshared coordination can maintain rapid rotation despite prolonged residence. An experimental examination of this effect could provide valuable insights into the viscosity, ionic conductivity, and thermal stability of LIBs.

\begin{acknowledgement}

RR acknowledges funding support from the Department of Science and Technology's National Supercomputing Mission grant DST/NSM/R\&D\_HPC\_Applications/2021/30. The authors acknowledge the Param Shivay (IIT BHU) and Param Ananta (IIT Gandhinagar) supercomputing facilities for the simulations reported in this work.

\end{acknowledgement}

\begin{suppinfo}

System size details, additional plots for radii of gyration, rotational autocorrelation functions, lifetimes, mean-squared displacements, validation of a charge-scaled force-field parameters are provided in the Supporting Information (Figures S1 - S11, Tables S1 - S6). 

\end{suppinfo}

\bibliography{Glyme-TFSI-paper}
\end{document}